\documentclass[fp,twocolumn]{jpsj3}
\usepackage{txfonts}
\usepackage{color}

\title{Search for Quantum-Tricritical-Point in Antiferromagnet CeRu$_2$(Si$_{1-x}$Ge$_x$)$_2$}

\author{Hiroki Shinya$^1$, Fumika Ito$^1$, Noriyuki Kabeya$^{1*}$, Yuta Mizukami$^1$, and Noriaki Kimura$^{1\dag}$}
\inst{$^1$Department of Physics, Graduate School of Science, Tohoku University, Sendai 980-8578, Japan} 

\abst{CeRu$_2$Si$_2$ is a well-known heavy fermion paramagnet, and substituting Ge for Si induces antiferromagnetism. This antiferromagnetism is Ising-like and has a tricritical point in the magnetic field ($H$) -temperature ($T$) phase diagram. Since the temperature of the tricritical point is expected to decrease with increasing pressure ($P$), we investigated the pressure dependence of the magnetic phase transitions. We determined the $H$-$P$ phase diagram and revealed that a first-order phase-transition line changes to a quantum critical line, which implies the existence of the quantum tricritical point. A ferromagnetic quantum fluctuation arises in the vicinity of the possible quantum tricritical point.
}

\begin{document}
\maketitle

\section{Introduction}
The magnetic quantum critical point (QCP) has long been the scene of interesting physics in strongly correlated electron systems, including the appearance of exotic superconductivity, non-Fermi liquids, and magnetic fluctuations\cite{herz1976,millis1993,moriya1985,sachdev2011,stewart2001,vojta1997,lohneysen2007}. In particular, many superconductors have been found near the QCP in heavy fermion antiferromagnets, and exploring new materials has been actively conducted\cite{pfleiderer2009}. On the other hand, from the viewpoint of quantum criticality, the magnetic phase diagram near the QCP has not yet been clarified experimentally because the induced superconductivity often masks the details of the critical phenomena.

This study focuses on the magnetic phase diagram of Ising-type antiferromagnets near the QCP. Ising-type antiferromagnets have a tricritical point (TCP) where the transition line changes from a second-order (continuous) to a first-order in the magnetic field ($H$) - temperature ($T$) phase diagram. The existence of TCP can be explained by the mean-field theory of the Ising model, which takes into account next-nearest neighbors. The Hamiltonian of the system is expressed as follows\cite{kincaid1974,stryjewski1977}.
\begin{eqnarray}
  \mathcal{H}=|J|\sum_{<nn>} S_{jz}S_{kz} + J'\sum_{<nnn>} S_{jz}S_{kz} -\mu H_i \sum_j S_{jz}.
  \label{eq:one}
\end{eqnarray}
Here, the first term on the right-hand side is the interaction among the nearest-neighbor pairs of spin ($\left<nn\right>$). The second term is the interaction among the next-nearest-neighbor pairs ($\left<nnn\right>$), where $S_{jz}$ is the $z$-component of the spin on site $j$, $\mu$ is the magnetic moment per spin, and $H_i$ is the internal magnetic field. When $R\equiv\frac{Z'J'}{Z|J|}<-\frac{3}{5}$, where $Z$ and $Z'$ are the numbers of nearest- and next-nearest-neighbors respectively, a TCP appears in the $H$-$T$ phase diagram. In other words, the TCP appears where the interaction among next-nearest-neighbors is ferromagnetic and competes with the antiferromagnetic interaction among nearest-neighbors. It is interesting to know whether there exists a quantum tricritical point (QTCP) where the TCP reaches absolute zero in such a situation, how QCTP differs from ordinary antiferromagnetic QCP, and what ground state is realized in the vicinity of the QTCP.

Misawa et al. theoretically discussed how a magnetic phase with TCP changes when the magnetism is suppressed by a control parameter, $g$, such as pressure\cite{misawa2008,misawa2009}. According to this, the metamagnetic transition field decreases with pressure at absolute zero, and at a specific pressure, the phase transition switches from first-order to continuous. The switching point is the QTCP. If pressure further increases, the metamagnetic transition becomes a continuous phase transition, and the transition field decreases as it is, eventually reaching zero magnetic field. The ground state becomes paramagnetic above this pressure. This model explains the field-induced QCP in YbRh$_2$Si$_2$ and other compounds\cite{misawa2008}. However, how the phase diagram shifts as pressure continuously changes is still unclear.

In order to elucidate the properties around the QTCP, we have selected a tetragonal ThCr$_2$Si$_2$-type crystal ($I4/mmm$), CeRu$_2$(Si$_{1-x}$Ge$_x$)$_2$, in which Si is substituted for Ge in CeRu$_2$Si$_2$. CeRu$_2$Si$_2$ is a well-known heavy fermion compound and is paramagnetic in its ground state but shows a metamagnetic crossover at 7.7~T\cite{haen1987}. The substitution of Ge works as a negative pressure and then induces Ising-type antiferromagnetism\cite{godart1986,mignot1991,haen1999}. The QCP at which the N\'{e}el temperature $T_{\rm N}$ drops to absolute zero is predicted to be $x_c=0.065$\cite{matsumoto2011}, but the details in the very vicinity of the QCP are not well understood because the parameter control by substitution does not allow sufficient resolution. In this study, we investigated the phase diagram of the $x=0.12$ sample, whose ground state is antiferromagnetic but close to the QCP, from the measurement of the magnetoresistance and Hall resistance. This compound exhibits a first-order metamagnetic transition at low temperatures\cite{sugi2006}, which is expected to have a TCP in the $H$-$T$ phase diagram. We further investigated the pressure dependence of the magnetic transition at low temperatures to search for the QTCP. The behavior of the $A$ coefficient of the electrical resistivity was also studied to evaluate the criticality.

\section{Experimental}
A single crystal with the nominal composition CeRu$_2$(Si$_{1-x}$Ge$_x$)$_2$, $x=0.12$, was grown by the Czochralski pulling method in a tetra-arc furnace with an Ar gas atmosphere. We have cut a rectangular sample ($a\times a\times c=2.5 \times 1.2 \times 0.25$mm$^3$) from the obtained single crystal ingot.
Hydrostatic pressures up to 1.06~GPa were produced by a clamped piston-cylinder cell. We used a $1:1$ mixture of 1- and 2-propanol as the pressure-transmitting medium. The pressures at low temperatures were determined by the resistivity of a Manganin wire that was calibrated against the drop of the ac susceptibility due to the superconducting transition of Sn.

Magnetoresistance and Hall resistance were measured with a conventional dc six-terminal method. The connection between the sample and the contact wire was made through spot welding. The electric current and magnetic field were applied along the $a$- and $c$-axes, respectively. The Hall voltage was derived by taking the difference between the values measured in the positive and negative magnetic fields. We utilized a dilution refrigerator with a 17~T superconducting magnet for the measurement below 2~K. We also used a PPMS (Quantum Design Co. Ltd.) and a variable temperature insert with a 14~T magnet for the measurements above 2~K at ambient pressure.

\section{Results and discussion}
\begin{figure}
\begin{center}
\includegraphics[width=\linewidth]{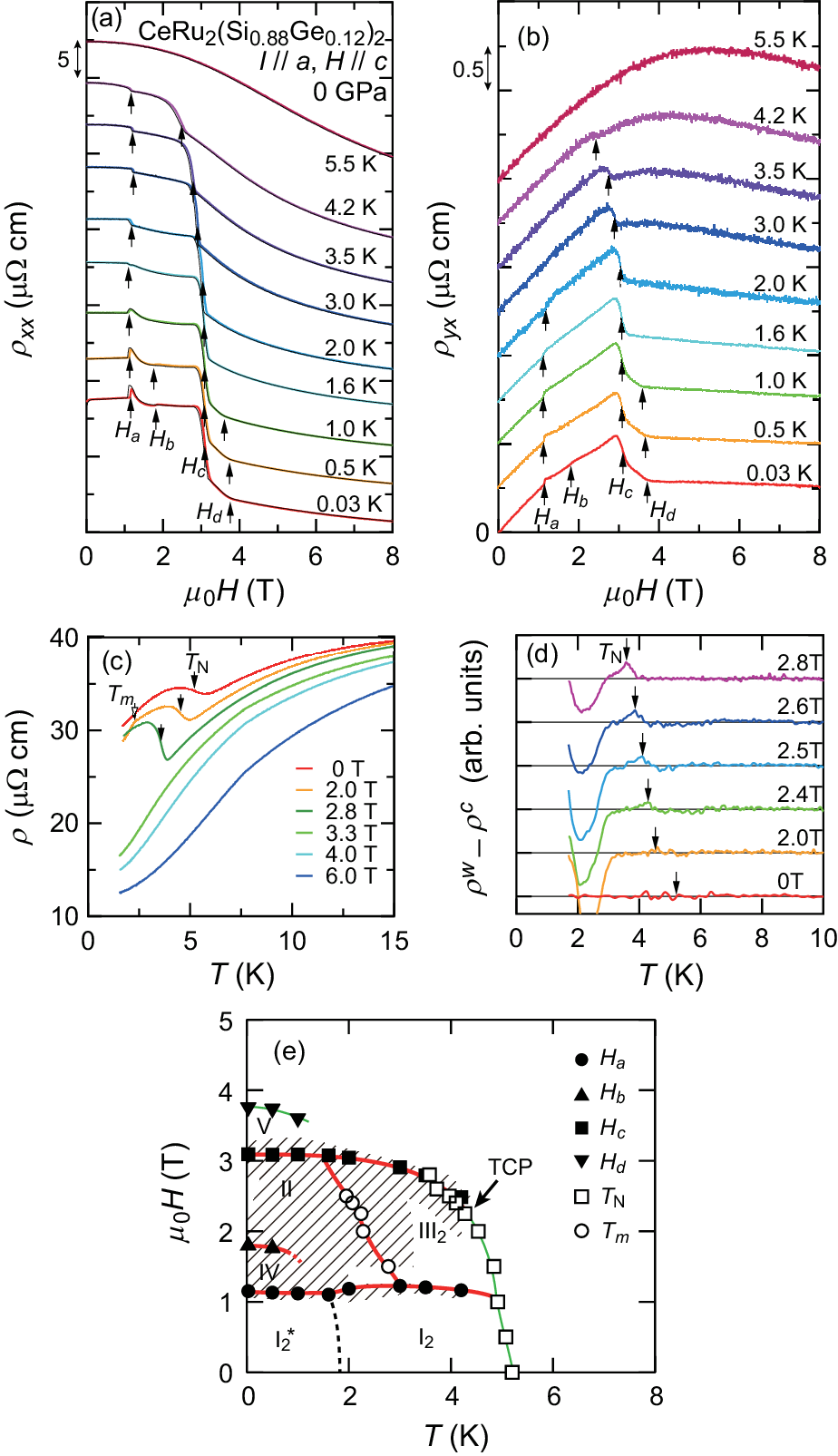}
\end{center}
\caption{(Color online) Temperature evolution of (a) Magnetoresistance $\rho_{xx}(H)$ and (b) Hall resistance $\rho_{yx}(H)$ of CeRu$_2$(Si$_{0.88}$Ge$_{0.12}$)$_2$ for $H\parallel c$ at ambient pressure. Each data is shifted vertically for clarity. The colored (gray) curves in (a) represent the field-raising process, while the black ones indicate the field-lowering process. The transition fields are indicated by arrows. Each transition field is almost the same in the magnetoresistance and Hall resistance. (c) Temperature dependence of the resistivity under magnetic fields. (d) Magnetic field variation of the difference in the resistivity between warming and cooling processes ($\rho^w$ and $\rho^c$). A finite difference indicates a hysteresis existing at a transition. (e) $H$-$T$ phase diagram constructed from (a) and (c). Hatching area means the hysteresis region. The red (thick) and green (thin) curves indicate the transition with and without hysteresis.
}
\label{fig:0GPa}
\end{figure}

To confirm the TCP, we first determined $H$-$T$ phase diagram at ambient pressure. We observed four anomalies in the magnetoresistivity $\rho_{xx}$ at the lowest temperature of 0.03~K, as shown in Fig.~\ref{fig:0GPa}(a). The Hall resistivity $\rho_{yx}$ also shows four anomalies in Fig.~\ref{fig:0GPa}(b). Here we define the inflection points of $\rho_{xx}(H)$ at the raising process of the magnetic field as the transition fields $H_a$, $H_b$, and $H_c$, and the bending point as $H_d$. Hysteresis was observed for all except for $H_d$ when the magnetic field was raised and lowered. We note that we do not plot the field-lowering process in the Hall resistivity because we can not derive correct hysteresis from the difference between the positive and negative magnetic fields. The anomalies at $H_b$ and $H_d$ faded as the temperature increased. The temperature dependence of electrical resistivity under the magnetic fields is shown in Fig.~\ref{fig:0GPa}(c). At 0~T, a hump in resistivity due to the antiferromagnetic transition is observed below 6~K. The inflection point is defined as the N\'{e}el temperature $T_{\rm N}=5.2$~K. At zero field, there was no hysteresis between warming and cooling processes, but hysteresis at $T_{\rm N}$ was observed above 2~T [Fig.~\ref{fig:0GPa}(d)]. Therefore, the TCP is estimated to be located near 2.3~T. Above 1~T, a bend in the resistivity with hysteresis was observed at temperatures ($T_m$) lower than $T_{\rm N}$ [Fig.~\ref{fig:0GPa}(c)]. Figure~\ref{fig:0GPa}(e) summarizes these results as a phase diagram. The observed hysteresis region is indicated by hatching. The transitions within this area are first-order ones. The first-order phase transition is indicated by the red line, and the transition without hysteresis is indicated by the green line. The obtained phase diagram is almost consistent with previous reports\cite{matsumoto2011}. The labels of the phases follow those in the literature. The phase boundary between I$_2$ and I$_2^*$ reported previously was not confirmed in this study because our measurement of the resistivity was limited to 1.7~K. $H_d$ is a newly observed anomaly. Since the anomaly disappears with increasing temperature, it is unclear whether the region between $H_c$ and $H_d$ is a distinct phase. In this study, however, we call this region phase V.
%

\begin{figure}
\begin{center}
\includegraphics[width=\linewidth]{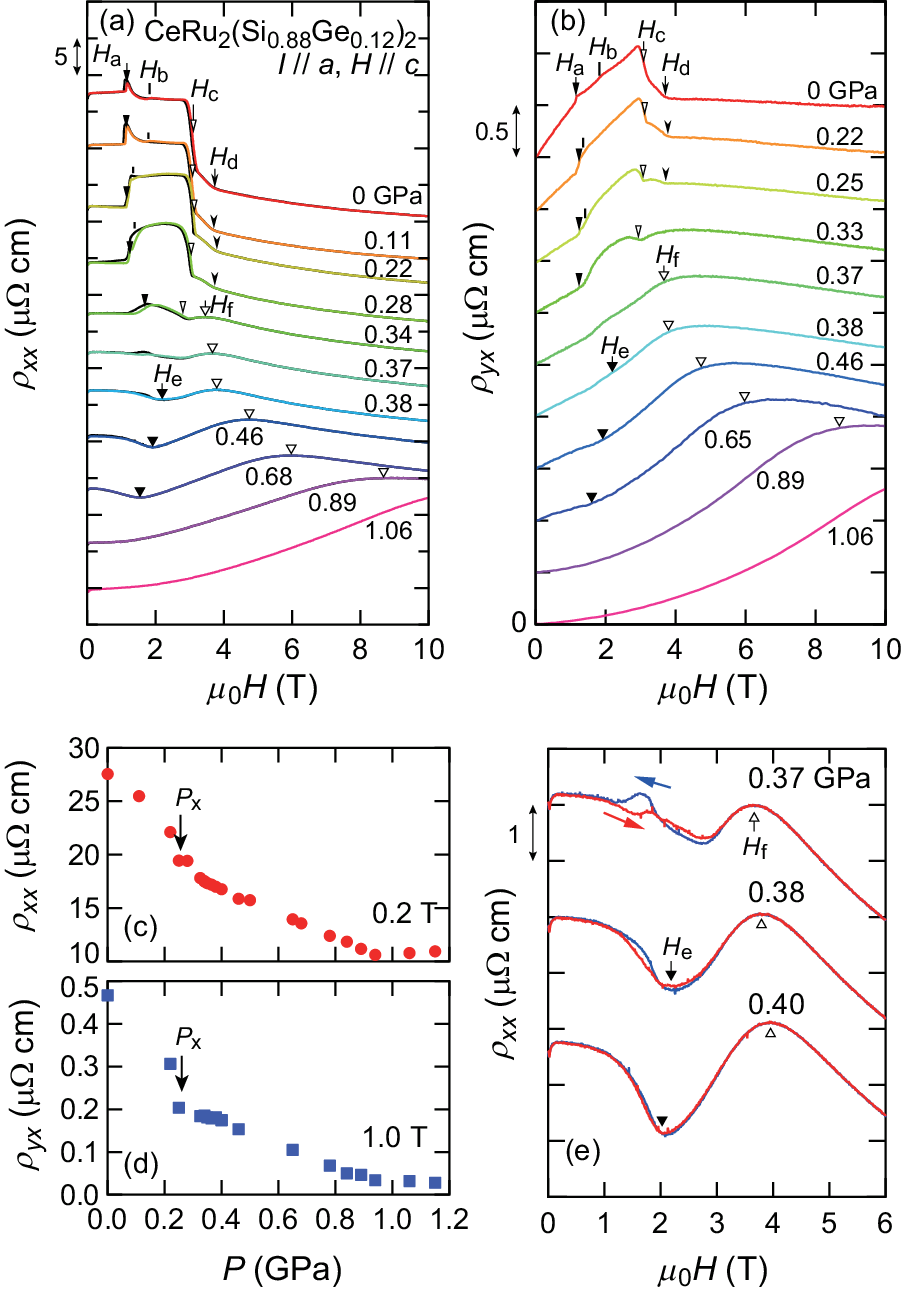}
\end{center}
\caption{(Color online) Pressure evolution of (a) Magnetoresistivity $\rho_{xx}(H)$ and (b) Hall resistivity $\rho_{yx}(H)$ at 0.03~K. Each data is shifted vertically for clarity. The colored curves in (a) represent the field-raising process, while the black ones indicate the field-lowering process. The transition fields are indicated by arrows. Since it was difficult to define the transition field from $\rho_{yx}(H)$, especially at higher pressure, for (b) we draw the arrows at the same fields as defined on $\rho_{xx}$ in (a). (c) Pressure dependence of $\rho_{xx}$ at 0.2~T. (d) Pressure dependence of $\rho_{yx}$ at 1.0~T. (e) The $\rho_{xx}(H)$-curves in raising- and lowering-field processes at pressures near the pressure at which hysteresis vanishes.
}
\label{fig:PdepMR}
\end{figure}

To search for the QTCP, we determined the $H$-$P$ phase diagram from the pressure dependence of $\rho_{xx}$ at the lowest temperature of 0.03~K. Figure~\ref{fig:PdepMR}(a) shows that the behavior of $\rho_{xx}$ changes significantly with the application of pressure. At 0~GPa, the transition at $H_a$ is accompanied by a peak, but it changes to a step increase with applying pressure. The $H_c$ transition is also rapidly obscured at pressures above 0.3~GPa and unclear at 0.37~GPa. Correspondingly, the anomaly of $H_d$ is obscured by the application of pressure. A slow peak appears to replace the disappearance of the inflection at $H_d$. The maximum field $H_f$ increases monotonically with the application of pressure. On the other hand, after the anomalies of $H_a$ and $H_c$ disappear, a new minimum in $\rho_{xx}$ appears from 0.38~GPa. The minimum field, $H_e$, shifts to the low field side with increasing pressure and falls to zero at approximately 0.9~GPa.
%

In the Hall resistivity $\rho_{yx}$, the anomalies at $H_a$, $H_b$, $H_c$, and $H_d$ become less obvious with increasing pressure. At 0.37~GPa, they are no longer identified clearly. The anomaly at $H_e$ appearing from 0.38~GPa is weaker than that in the magnetoreistivity. $H_f$ corresponds to where $\rho_{yx}(H)$-curve changes in slope. Since the metamagnetic crossover appears after the disappearance of the magnetic transitions at $H_a$, $H_b$, and $H_c$ in the Ge-substitution system\cite{sugi2006,sugi2008,matsumoto2011}, $H_f$ is expected to correspond to the metamagnetic crossover field. The peak of the megnetoresistivity and change in slope of the Hall resistivity are the characteristic features of the metamagnetic crossover\cite{haen1987}.

The resistivity and Hall resistivity on the low field side below $H_a$ changes significantly with the application of pressure, as shown in Figs.~\ref{fig:PdepMR}(c) and (d). A rapid change is observed at 0.25~GPa. This suggests that the phase I$_2^*$ undergoes a phase transition at 0.25~GPa ($\equiv P_x$). We speculate that this higher-pressure state is phase I$_2$ or III$_2$ which is located on the high-temperature side of the phase I$_2^*$ or II in the $H$-$T$ phase diagram at ambient pressure. Both the resistivity and Hall resistivity show a minimum of around 1.0~GPa, corresponding to the pressure at which $H_e$ falls to 0~T.

\begin{figure}
\begin{center}
\includegraphics[width=\linewidth]{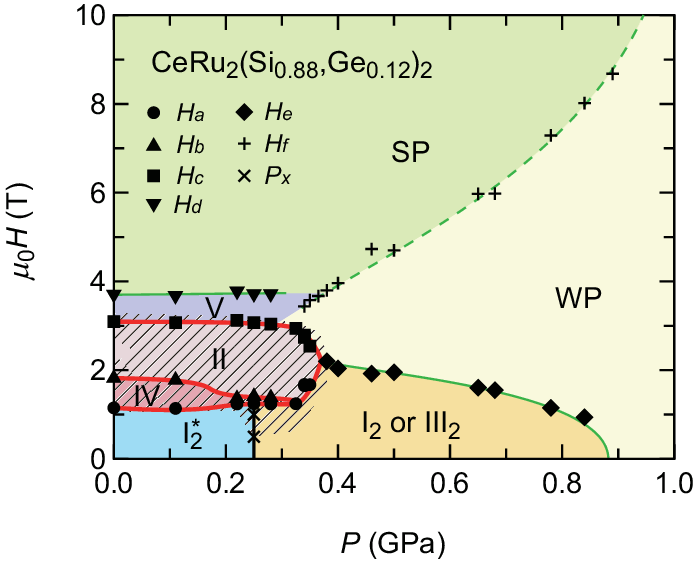}
\end{center}
\caption{(Color online) $H$-$P$ phase diagram at 0.03~K in CeRu$_2$(Si$_{0.88}$Ge$_{0.12}$)$_2$. The red (thick) and green (thin) curves indicate the transition with and without hysteresis. The black line represents the transition where the hysteresis cannot be confirmed. The dashed curve is a possible metamagnetic crossover. Hatching area means the hysteresis region.
}
\label{fig:H-P}
\end{figure}

From the pressure dependence of the transition fields in conjunction with the resistivity and Hall resistivity, we constructed a $H$-$P$ phase diagram at the base temperature, as in Fig.~\ref{fig:H-P}. Up to 0.3~GPa, the values of $H_a$, $H_c$, and $H_d$ are robust, but from 0.3~GPa to 0.4~GPa, $H_a$ and $H_c$ rapidly approach each other. The transition field $H_b$ merges into the $H_a(P)$-line at 3.4~GPa. The transition fields $H_a$ and $H_c$ meet together and then phase II closes at 0.38 GPa. The $H_e(P)$-line appears from the closing point to the higher pressure side. The $H_f(P)$-line also emerges from 0.3~GPa. If $H_f$ is the metamagnetic crossover field, we consider the low-field side below $H_f$ to be a weakly polarized (WP) phase and the high-field side to be a strongly polarized (SP) phase. Since the $H_d$ line becomes obscure with increasing pressure, it is unclear how phase V closes around 0.3~GPa.

Next, we discuss the location of the QTCP on the $H$-$P$ phase diagram. Since the TCP exists on the $H_c$ transition line at ambient pressure, as shown in Fig.~\ref{fig:0GPa}(e), we trace the order of phase transition in the $H_c(P)$-line in Fig.~\ref{fig:H-P}. As shown in Fig.~\ref{fig:PdepMR}(e), the hysteresis in the magnetoresistivity was observed up to 0.38~GPa, which is the emerging point of the $H_e(P)$-line. At 0.4~GPa, the hysteresis was not observed. Therefore, we estimate the QTCP is located at the emerging point or slightly higher. We cannot conclude that the emerging point of the $H_e(P)$-line corresponds to the QTCP within our experimental accuracy. 

It is critical to determine if the QTCP corresponds to the emerging point of the $H_e(P)$ line. When this point is the QTCP, it is strange that the two first-order phase-transition lines, namely, $H_a(P)$- and $H_c(P)$-lines, emerge from the QTCP, because a QTCP should be a point at which a single first-order phase transition line switches to a second-order one. Therefore, the $H_e(P)$-line should be a first-order at least just above the emerging pressure of $H_e$, that is, the emerging point of $H_e$ should be a triple point. From this consideration, the QTCP may be located on the slightly high-pressure side of the triple point.More precise location of the QTCP could be clarified through a detailed study of the pressure dependence of the $H$-$T$ phase diagram.

\begin{figure}
\begin{center}
\includegraphics[width=\linewidth]{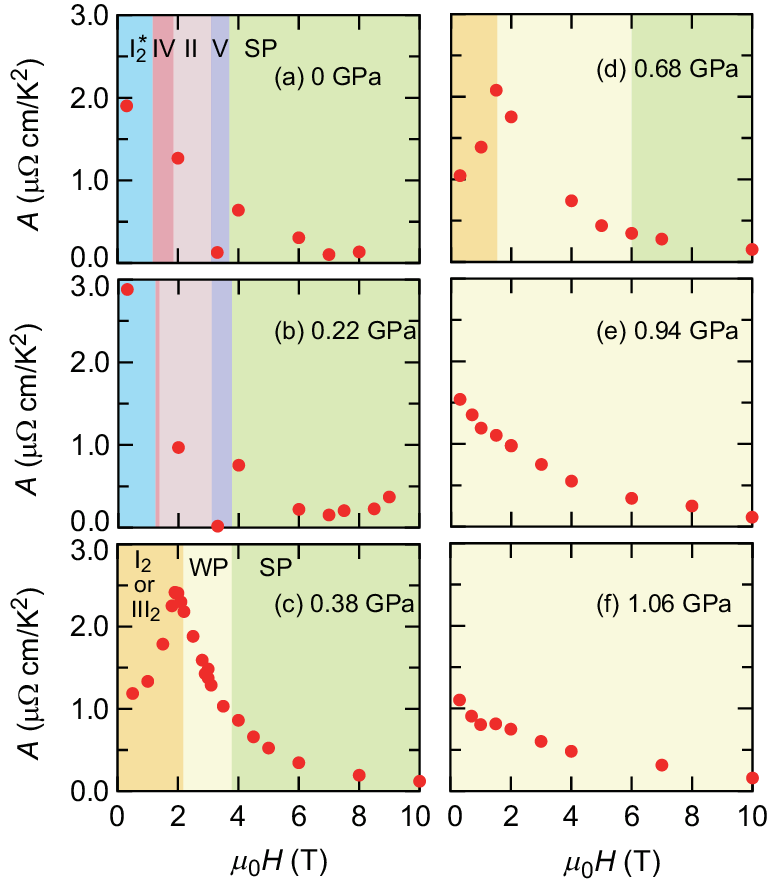}
\end{center}
\caption{(Color online) Pressure evolution of the $A$ coefficient as a function of the magnetic field. The magnetic phases are indicated by different colors (intensities of gray).
}
\label{fig:Acoef}
\end{figure}

It is expected to be a quantum critical line (QCL) at pressures above the QTCP. To investigate the criticality, we measured the temperature dependence of the electrical resistivity and estimated the $A$ coefficient where $\rho=\rho_0+AT^2$ by fitting the data in the temperature range below 0.3~K. The magnetic field dependence of the $A$ coefficient at various pressures is shown in Fig.~\ref{fig:Acoef}. At the lower pressures below the QTCP, the $A$ coefficient tends to decrease toward higher magnetic fields [Figs.~\ref{fig:Acoef}(a) and (b)]. In phase V, however, the $A$ coefficient was exceptionally small. At 0.38~GPa, the $A$ coefficient shows a divergent behavior toward $H_e$, the boundary between phase I$_2$ or III$_2$ and WP state [Fig.~\ref{fig:Acoef}(c)]. A similar peak structure is observed at higher pressures of 0.68~GPa, but the maximum value is smaller than that at 0.38~GPa [Fig.~\ref{fig:Acoef}(d)]. These indicate that the $H_e(P)$-line is a QCL that exhibits quantum criticality.
%

When we look at the $A$ coefficient variation at the lowest magnetic field in Figs.~\ref{fig:Acoef}(a) to (c), we realize it takes the maximum at 0.22~GPa. Thus, the phase boundary between phase I$_2^*$ and phase I$_2$ or III$_2$ at 0.25~GPa can also be a QCL where the electronic correlation is enhanced.

\begin{figure}
\begin{center}
\includegraphics[width=\linewidth]{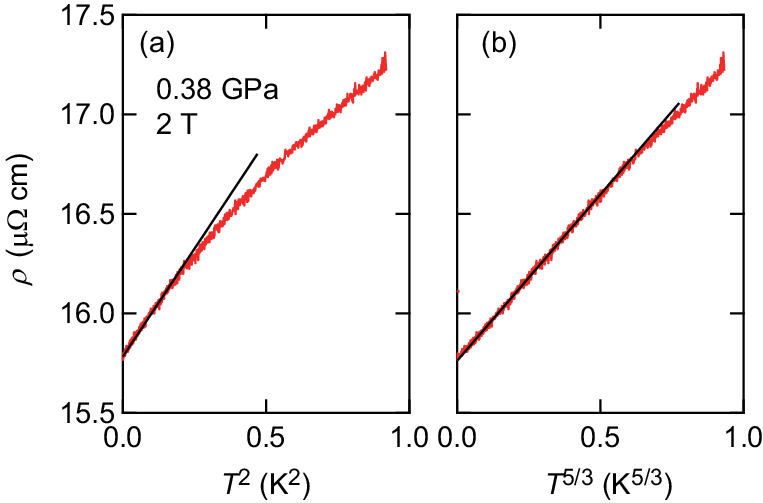}
\end{center}
\caption{(Color online) (a) $T^2$ and (b) $T^{5/3}$ plots of the resistivity at 0.38~GPa and 2~T close to the QTCP.
}
\label{fig:Power}
\end{figure}

In order to estimate the criticality in the vicinity of the possible QTCP, we compared the $T^2$ and $T^{5/3}$-plots of the resistivity at 0.38~GPa, as shown in Fig.\ref{fig:Power}. The resistivity follows well $T^{5/3}$ rather than $T^2$, suggesting that the 3D ferromagnetic fluctuation dominates electron scattering\cite{moriya1985}. Naively, it does not seem surprising that the ferromagnetic fluctuation plays a vital role in the QTCP because the exchange interaction between next-nearest-neighbors must be ferromagnetic for the TCP to appear, as mentioned in Eq.~(\ref{eq:one}). In fact, since ferromagnetism appears at higher Ge concentrations in Ge-substituted systems\cite{haen1996,haen1999,sugi2008}, ferromagnetic interactions can survive even at lower concentrations.
%

According to the theory of QTCP by Misawa et al.\cite{misawa2009}, the temperature dependence of the resistivity at the QTCP is expected to be similar to that of the conventional QCP if only a relaxation time contributes to the resistivity. In the case of the present compound, $T^{3/2}$ for 3D antiferromagnetic fluctuation is expected\cite{moriya1985}. However, the theory also pointed out that if a change in the carrier density accompanies the quantum criticality, the temperature dependence is more complicated\cite{misawa2009}. Further verification of the consistency with theory is considered necessary. From an experimental point of view, verification from various aspects such as magnetic susceptibility, specific heat, and $T_1$ is needed.

Finally, we refer to the metamagnetic crossover. As mentioned earlier, $H_f$ is considered to correspond to a metamagnetic crossover. However, as shown in Fig.~\ref{fig:Acoef}, no enhancement toward $H_f$ in the $A$ coefficient was observed. This is in contrast to the results in CeRu$_2$Si$_2$\cite{kambe1995} and Ce(Ru,Rh)$_2$Si$_2$\cite{aoki2012}. The reason for this was not apparent. It may be a peculiar property of the Ge-substitution system.

\section{Summary}
This study investigated the $H$-$P$ phase diagram below 0.1~K in the Ising antiferromagnet CeRu$_2$(Si$_{0.88}$Ge$_{0.12}$)$_2$ from the magnetoresistivity and Hall resistivity. We then determined the phase diagram. From the pressure dependence of the hysteresis in the magnetoresistivity, we estimate the QTCP to be located at or slightly higher than the pressure of the closing point of phase II. To obtain conclusive evidence of the QTCP, more detailed measurements are required. Furthermore, we found the QCL that arises from the possible QTCP. The QCL dropped to zero field with the application of pressure. The phase boundary between the phase I$_2^*$ and higher phase (I$_2$ or III$_2$) is possibly another QCL. The obtained phase diagram is more complicated than the theoretically predicted phase diagram. The $A$ coefficient of electrical resistivity was observed along the QCLs. The temperature power of the resistivity at the possible QTCP is $5/3$, implying the ferromagnetic fluctuation arising. We did not observe an enhancement of the $A$ coefficient along the metamagnetic crossover.
\vspace{5mm}
\begin{acknowledgment}
We thank M. Kikuchi, H. Moriyama, and Y. Shimakoshi for their technical support. This work was supported by JSPS KAKENHI Grant Numbers JP24K00585, JP23H04871, JP23K03332, JP23K20823, JP23K25829, JP22H00102, JP22K03505, JP22740241, and JP19740222. This research was also supported by JST SPRING, Grant Number JPMJSP2114 and by the 11th Research Grant of the Hirose Foundation.
\end{acknowledgment}

\vspace{5mm}
\noindent
$^\dag$noriaki.kimura.a6@tohoku.ac.jp\\
Present address: \\
$^{*}$Faculty of Science, University of Ryukyus, Okinawa 903-0213, Japan\\

\bibliographystyle{jpsj}
\bibliography{ceru2sige2bib}

\end{document}